\documentclass[12pt]{article}

\normalbaselines \topmargin -0.5 truein  \oddsidemargin 0.30
truein \evensidemargin 0.30 truein 
scaled \magstep2  \font\tbf = cmbx12

\begin{document}


\indent \vskip 1cm \centerline{\tbf  NON-MINIMAL COUPLING OF
PHOTONS AND AXIONS}

\vskip 0.8cm


 \vskip 0.3cm \centerline{\tbf Alexander B.
Balakin\footnote{e-mail: Alexander.Balakin@ksu.ru} } \vskip 0.3cm
\centerline{\it Kazan State University, Kremlevskaya street 18,
420008, Kazan,  Russia,} \vskip 0.5cm \centerline{and} \vskip
0.5cm \centerline{\tbf Wei-Tou Ni\footnote{e-mail: wtni@pmo.ac.cn}
} \vskip 0.3cm \centerline{\it Center for Gravitation and
Cosmology, Purple Mountain Observatory,} \centerline{\it Chinese
Academy of Sciences, Nanjing, 210008, People's Republic of China, and} \centerline{\it
National Astronomical Observatories, Chinese Academy of Sciences,}
\centerline{\it
Beijing, 100006,  People's Republic of China} \vskip 3cm

\vskip 3cm

{\tbf Abstract} \quad {\small We establish a new self-consistent
system of equations accounting for a non-minimal interaction of
gravitational, electromagnetic and axion fields. The procedure is
based on a non-minimal extension of the standard
Einstein-Maxwell-axion action. The general properties of a
ten-parameter family of non-minimal linear models are discussed.
We apply this theory to the models with pp-wave symmetry and
consider propagation of electromagnetic waves non-minimally
coupled to the gravitational and axion fields. We focus on exact
solutions of electrodynamic equations, which describe
quasi-minimal and non-minimal optical activity induced by the axion
field. We also discuss empirical constraints on coupling
parameters from astrophysical birefringence and polarization
rotation observations.}

\vskip 0.5cm

PACS numbers: 04.20.-q, 04.40.-b, 04.40.Nr

Key words: Non-minimal interactions, axion field, birefringence,
polarization rotation

\newpage

\section{Introduction}

The topic of pseudoscalar-photon interaction and axion  theory is
a very attractive branch of modern physics, and there are two
reasons for this interest. First of all, it is an interest to
axions as hypothetical particles, which appear in the context of
the strong CP-violation problem and spontaneous breaking of
symmetry in the early universe (see, e.g., \cite{Kallosh,Peccei2}
for review and basic references), and which are considered as one
of the Dark Matter candidates (see, e.g., \cite{Raffelt,Turner}
and references therein). The direct detection of the axions is
formulated to be one of goals of modern experiments in High Energy
Physics (see, e.g., \cite{Battesti}).

The second reason is connected with an interest to the theoretical
concept of pseudoscalar field $\phi$ (axion field) associated with
pseudoscalar-photon interaction. This occurs first in the study of
electromagnetism and equivalence principles \cite{6Ni,7Ni,8Ni}.
This concept is a base for the so-called Einstein-Maxwell-axion
theory, Einstein-Maxwell-dilaton-axion theory  and their
generalizations (see, e.g., \cite{6Ni}-\cite{EMA5}).

One of the most significant item in the theory of axion fields is
the problem of interaction of electromagnetic and pseudoscalar
(axion) fields, the well-known application of the photon-axion
coupling being the effect of polarization rotation
\cite{6Ni,rot2,rot3} in the observations of Cosmic Microwave
Background (CMB) radiation (see, e.g., \cite{19,20} and references
therein). Several groups are working on the experiments with
vacuum birefringence and vacuum dichroism to look for the
photon-axion coupling \cite{21}-\cite{24}.

It is well-known that the standard effect of optical activity
induced by the axion field takes place when the pseudoscalar field
$\phi$ has a non-vanishing four-gradient \cite{8Ni,HehlObukhov},
i.e. when $\partial_k \phi \neq 0$. This effect can be described
in the framework of classical electrodynamics in terms of linear
constitutive equations \cite{8Ni,HehlObukhov,HO3}, linking an
excitation tensor $H^{ik}$ and the Maxwell tensor $F_{mn}$. When
the pseudo-scalar field is constant, $\phi=\phi_0$, and thus does
not contribute to the vacuum Maxwell equations, new possibilities
for the axion-photon interaction exist: one of them is discussed
in the work \cite{HO2}, which focuses on reflection and
transmission of a wave at an interface between two media. We
discuss here another possibility related to {\it non-minimal}
optical activity induced by photon-axion coupling, which also can
occur in the case of a constant pseudoscalar field.

The standard description of the non-minimal coupling of the
gravitational field with scalar, electromagnetic and gauge fields
is based on the introduction of specific cross-terms into the
Lagrangian, which contain the Riemann tensor $R_{ikmn}$, the Ricci
tensor, $R_{kn}$, and Ricci scalar, $R$, on the one hand, and the
corresponding fields and their derivatives, on the other hand. The
theory of non-minimal coupling is elaborated in detail for the
scalar field (real and complex fields, as well as Higgs
multiplets)(see, e.g., \cite{FaraR} for review and references).
Special attention in these investigations is focused on two
models, the first of them has the $\xi R \Phi^2$ coupling, and the
second has the so-called non-minimal derivative coupling
\cite{derc1}-\cite{derc3}. The study of a non-minimal coupling of
gravity with electromagnetic field started in \cite{Prasanna1} and
this theory has been developed by many authors (see, e.g.,
\cite{HO}-\cite{BBL08}). The generalization of the idea of
non-minimal interactions to the case of torsion coupled to the
electromagnetic field has been made in \cite{Hehl1,Hehl2}. Exact
solutions in the framework of non-minimal Einstein-Yang-Mills and
Einstein-Yang-Mills-Higgs theories are discussed in
\cite{BZ1}-\cite{BDZ3}.

In this paper we focus on the non-minimal generalization of the
Einstein-Maxwell-axion theory. The derivation of master equations
is fulfilled in analogy with non-minimal scalar field theory and
non-minimal electromagnetic field theory. The items which are
specific for the pseudoscalar field theory are discussed in
detail. The paper is organized as follows. In section 2, we discuss
a formalism of non-minimal Einstein-Maxwell-axion theory; a brief
introduction to the minimal theory is given in subsection 2.1; the
non-minimal extension of the Lagrangian and the derivation of the
corresponding master equations are presented in subsection 2.2;
subsection 2.3 contains the description of the model in terms of
constitutive equations. In section 3 we apply the obtained master
equations to the models with pp-wave symmetry and discuss examples
of the exact solutions for the gravitational, pseudoscalar and
electromagnetic fields. In section 4 we consider the
propagation of test electromagnetic waves coupled to pseudoscalar
field in the pp-wave background and focus on the effect of
non-minimal optical activity induced by the axion field. In section 5
we summarize the results.

\section{Non-minimal coupling of gravitational, \\ electromagnetic and axion fields}

\subsection{Minimal model as a starting point}

In order to explain the novelty of our approach, let us first
introduce the case of gravitational-electromagnetic-axion fields
minimally coupled. The simplest minimal action functional is
\begin{equation}
S_{({\rm M})} {=} \int d^4 x \sqrt{{-}g} \left[\frac{R}{\kappa}
{+} \frac{1}{2} F^{mn}F_{mn} {+} \frac{1}{2} \phi F^{*mn}F_{mn}
{-} g^{mn} \nabla_m \phi \nabla_n \phi {+} m^2_{({\rm A})} \phi^2
\right] \,, \label{actmin}
\end{equation}
where $g$ is the determinant of the metric tensor $g_{ik}$,
$\nabla_{m}$ is a covariant derivative, $R$ is the Ricci scalar.
The Maxwell tensor $F_{mn}$ is given by
\begin{equation}
F_{mn} \equiv \nabla_m A_{n} - \nabla_n A_{m} \,,
\label{maxtensor}
\end{equation}
where $A_m$ is an electromagnetic potential four-vector; $F^{*mn}
\equiv \frac{1}{2} \epsilon^{mnpq}F_{pq}$ is the tensor dual to
$F_{pq}$; $\epsilon^{mnpq} \equiv \frac{1}{\sqrt{-g}} E^{mnpq}$ is
the Levi-Civita tensor, $E^{mnpq}$ is the absolutely antisymmetric
Levi-Civita symbol with $E^{0123}=1$. The dual Maxwell tensor
satisfies the condition
\begin{equation}
\nabla_{k} F^{*ik} =0 \,. \label{Emaxstar}
\end{equation}
The first term in the brackets is the Hilbert-Einstein Lagrangian;
the second term is the standard Lagrangian for an electromagnetic
field; the third term is the pseudoscalar-photon interaction
Lagrangian \cite{6Ni,7Ni,8Ni}; the fourth and fifth terms
constitute the pseudoscalar Lagrangian.

The symbol $\phi$ stands for a pseudoscalar field; this quantity
is dimensionless providing the terms $\frac{1}{2} F^{mn}F_{mn}$
and $ \frac{1}{2} \phi F^{*mn}F_{mn}$ to have the same
dimensionality. The axion field itself, $\Phi$, is considered to
be proportional to this quantity $\Phi = \Psi_0 \phi$ with a
constant $\Psi_0$. In order to simplify the calculations we
replace the potential of electromagnetic field $A_i$ with the term
$A_i / \Psi_0$, and consider the constant $\kappa$ to be equal to
$\kappa = \frac{8 \pi G \Psi^2_0}{c^4}$. The term $m_{({\rm A})}$
is proportional to a (hypothetical) mass of an axion, $m_{({\rm
A})} = 2\pi c \ m_{({\rm axion})} / h$; $h$ is the Planck
constant. We use the signature $+---$.

The variation of the action functional (\ref{actmin}) with respect
to the four-vector potential $A_i$ gives the minimal (vacuum)
Maxwell equations
\begin{equation}
\nabla_{k} \left[F^{ik} +  \phi F^{*ik} \right] = 0 \,,
\label{Emaxmin}
\end{equation}
which can be transformed into
\begin{equation} \nabla_{k} F^{ik} +
\ F^{*ik} \nabla_k \phi = 0  \label{Emaxmin1}
\end{equation}
due to the equation (\ref{Emaxstar}). Equivalently, equation
(\ref{Emaxmin}) can be written as
\begin{equation}
\nabla_{k} H^{ik} = 0 \,, \quad H^{ik} = C^{ikmn}F_{mn} \,,
\label{induc1}
\end{equation}
where $C^{ikmn}$ is the constitutive tensor
\begin{equation}
C^{ikmn} = \frac{1}{2} \left(g^{im} g^{kn} - g^{in}g^{km} \right)
+ \frac{1}{2} \phi \epsilon^{ikmn} \,, \label{induc2}
\end{equation}
and $H^{ik}$ is the excitation tensor \cite{8Ni,HehlObukhov}.
Minimal equations for the axion field  can be obtained from the
action (\ref{actmin}) by the variation with respect to the
pseudoscalar field $\phi$, yielding
\begin{equation}
\left[\nabla^k  \nabla_k {+} m^2_{({\rm A})} \right] \phi = -
\frac{1}{4} F^{*mn}F_{mn}  \,. \label{induc3}
\end{equation}
Minimal equations for the gravitational field obtained by the
variation of (\ref{actmin}) with respect to metric $g^{ik}$ are
\begin{equation}
R_{ik} - \frac{1}{2}Rg_{ik} = \kappa \left[ T^{({\rm EM})}_{ik} +
T^{({\rm A})}_{ik} \right] \,. \label{EineqMIN}
\end{equation}
Here $T^{({\rm EM})}_{ik}$ is the standard stress-energy tensor of
pure electromagnetic field:
\begin{equation}
T^{({\rm EM})}_{ik} \equiv  \frac{1}{4}g_{ik} F^{mn}F_{mn} -
F_{im}F_{k}^{ \ m} \,, \label{TEM}
\end{equation}
and $T^{({\rm A})}_{ik}$, given by
\begin{equation}
T^{({\rm A})}_{ik} \equiv \nabla_i \phi \nabla_k \phi -
\frac{1}{2} g_{ik} \left[ \nabla^m \phi \nabla_m \phi -
m^2_{({\rm A})} \phi^2 \right] \,, \label{TAX}
\end{equation}
is the stress-energy tensor of the pseudoscalar (axion) field
$\phi$.

\subsection{Non-minimal extension of the model}

\subsubsection{Lagrangian and non-minimal susceptibilities}

We consider the total action functional as a sum of minimal and
non-minimal contributions:
\begin{equation}
S=S_{({\rm M})} + S_{({\rm NM})} \,, \label{plus}
\end{equation}
where $S_{({\rm M})}$ is given by (\ref{actmin}). The non-minimal
contribution $S_{({\rm NM})}$ is, generally, a nonlinear function
of all independent invariants containing Riemann tensor $R_{iklm}$
and the dual ones $^{*}R_{iklm}$, $R^{\ *}_{iklm}$, Ricci tensor
$R_{mn}$ and Ricci scalar $R$ in convolutions with the Maxwell tensor
and its dual, as well as with derivatives of pseudoscalar field.
The set of such invariants for pure electromagnetic field was
discussed in \cite{BL05}; then it was supplemented by the
invariants containing scalar Higgs fields in
\cite{BDZ1,BDZ2,BDZ3}. For the axion field, this set of invariants
can be constructed analogously, and here we focus only on the
following Lagrangian linear in the curvature:
\begin{equation}
S_{({\rm NM})} {=} \int d^4 x \sqrt{{-}g} \left\{ \frac{1}{2}
{\cal R}^{ikmn} F_{ik}F_{mn} {+}  \frac{1}{2} {\chi}^{ikmn}_{({\rm
A})}  \phi \ F_{ik} F^{*}_{mn} {-} \Re^{mn}_{({\rm A})} \nabla_m
\phi \nabla_n \phi {+} \eta_{({\rm A})} R  \phi^2 \right\} \,.
\label{actnm}
\end{equation}
The quantity ${\cal R}^{ikmn}$ is a non-minimal three-parameter
susceptibility tensor \cite{BL05}, which has a form
\begin{equation}
{\cal R}^{ikmn} =  q_1 R g^{ikmn} + q_2 \Re^{ikmn} + q_3 R^{ikmn}
\,, \label{sus1}
\end{equation}
where
\begin{equation}
g^{ikmn} \equiv \frac{1}{2}(g^{im}g^{kn} {-} g^{in}g^{km}) \,,
\label{rrr}
\end{equation}
\begin{equation}
\Re^{ikmn} \equiv \frac{1}{2} (R^{im}g^{kn} {-} R^{in}g^{km} {+}
R^{kn}g^{im} {-} R^{km}g^{in}) \,. \label{rrrr}
\end{equation}
The constants $q_1$, $q_2$ and $q_3$ are non-minimal parameters
describing the linear coupling of the Maxwell tensor $F_{mn}$ with
curvature \cite{BL05}. The quantity ${\chi}^{ikmn}_{({\rm A})}$
given by
\begin{equation}
{\chi}^{ikmn}_{({\rm A})} {=} Q_1 R g^{ikmn} {+} Q_2\Re^{ikmn}
{+} Q_3 R^{ikmn} \,, \label{sus2}
\end{equation}
where $Q_1$, $Q_2$ and $Q_3$ are also constants, is the
non-minimal susceptibility tensor describing the linear coupling
of the dual tensor $F^{*}_{mn}$ with curvature. As in the previous
case, the combination $\phi F^{*}_{mn}$ gives the tensor quantity.
The tensor
\begin{equation}
\Re^{mn}_{({\rm A})} \equiv \frac{1}{2} \eta_1
\left(F^{ml}R^{n}_{\ l} + F^{nl}R^{m}_{\ l} \right) + \eta_2 R
g^{mn} + \eta_3 R^{mn} \label{sus3}
\end{equation}
describes a non-minimal susceptibility for the pseudoscalar field
in analogy with \cite{BDZ1}, but in this case it contains a term
linear in the Maxwell tensor. This term describes effects
analogous to the so-called derivative coupling in the non-minimal
scalar field theory \cite{derc1,derc2,derc3}. As for the tenth
coupling constant $\eta_{({\rm A})}$, it is a direct analog of the
well-known coupling constant $\xi$ in the non-minimal scalar field
theory (see, e.g., \cite{FaraR} for review and references).

The tensors ${\cal R}^{ikmn}$ and  ${\chi}^{ikmn}_{({\rm A})}$,
defined by (\ref{sus1}) and (\ref{sus2}), are skew-symmetric with
respect to transposition of the indices $i$ and $k$, as well as
$m$ and $n$. In addition, the following relations take place
\begin{equation}
{\cal R}^{ikmn}= {\cal R}^{mnik} \,, \quad {\chi}^{ikmn}_{({\rm
A})} = {\chi}^{mnik}_{({\rm A})}\,, \label{sus8}
\end{equation}
which guarantee that the model under consideration does not
contain the solutions of the skewon type \cite{HehlObukhov}. The
tensor $\Re^{mn}_{({\rm A})}$ is explicitly symmetric, i.e.
$\Re^{mn}_{({\rm A})} = \Re^{nm}_{({\rm A})}$.

\subsubsection{Non-minimal electrodynamic equations}

Electrodynamic equations, which correspond to the Lagrangian
(\ref{plus}) with (\ref{actmin}) and (\ref{actnm}), are linear and have the standard
form
\begin{equation}
\nabla_k  H^{ik} = I^i \,. \label{max2}
\end{equation}
The excitation tensor $H^{ik}$ and the Maxwell tensor $F_{mn}$ are
linked by the linear constitutive law
\begin{equation}
H^{ik} \equiv F^{ik} + {\cal R}^{ikmn} F_{mn} + \left[\phi \left(
F^{*ik} + {\chi}^{ikmn}_{({\rm A})} F^{*}_{mn} \right)\right] \,.
\label{inducnm}
\end{equation}
The second term on the right-hand side is the curvature-induced
polarization-magnetization, appearing in the non-minimally
extended pure Einstein-Maxwell model \cite{BL05}. The contribution
detailed in square brackets is the one of the axion type, the
terms $\phi$ and $F^{*ik}$ enter the equations in the
multiplicative form only (see, e.g., \cite{8Ni,HO} for more
details). The non-minimal axion contribution $ \phi
{\chi}^{ikmn}_{({\rm A})} F^{*}_{mn}$ is a new term proposed here.

The four-vector of an effective electric current
\begin{equation}
I^i \equiv \frac{1}{2}\eta_1 \nabla_k \left[ \left(R^{km} \nabla^i
\phi - R^{im} \nabla^k \phi \right) \nabla_m \phi \right]
\label{current}
\end{equation}
is proportional to the coupling parameter $\eta_1$ and is due to
the first term in (\ref{sus3}). This four-vector does not contain
Maxwell tensor and satisfies the conservation law $\nabla_i I^i =
0$. It contains both the Ricci tensor and the pseudoscalar field;
thus, it describes the electric current induced by curvature and
axion field derivatives.

\subsubsection{Non-minimal equation for the pseudoscalar field}

Non-minimally extended master equation for the pseudoscalar $\phi$
takes the form
\begin{equation}
\nabla_m \left[ \left( g^{mn} + \Re^{mn}_{({\rm A})} \right)
\nabla_n \phi \right] + \left[m^2_{({\rm A })} + \eta_{({\rm A})}
R \right] \phi = - \frac{1}{4} F^{mn}F^{*}_{mn} - \frac{1}{4}
{\chi}^{ikmn}_{({\rm A})} \ F_{ik} F^{*}_{mn} \,, \label{eqaxi1}
\end{equation}
where $\Re^{mn}_{({\rm A})}$ and ${\chi}^{ikmn}_{({\rm A})}$ are
given by (\ref{sus3}) and (\ref{sus2}), respectively. This
equation is a non-minimal generalization of (\ref{induc3}).

\subsubsection{Non-minimal equations for the gravitational field}

Variation of the action functional (\ref{plus}) with
(\ref{actmin}) and (\ref{actnm}) with respect to $g^{ik}$ gives
the non-minimally extended equations for the gravitational field
\begin{equation}
\left(R_{ik}{-}\frac{1}{2}Rg_{ik}\right) \left(1{+}\kappa \Theta
\right) {=} \kappa \left[  T^{({\rm EM})}_{ik}  {+} T^{({\rm
A})}_{ik} {+} T^{({\rm NMEM})}_{ik} {+} {\cal T}^{({\rm
NMA})}_{ik} \right] \,. \label{Eineq}
\end{equation}
Here the scalar $\Theta$ stands for the quantity
\begin{equation}
\Theta \equiv \eta_{({\rm A})} \phi^2 {+} \frac{1}{2} q_1
F_{mn}F^{mn} {+} \frac{1}{2} Q_1 \phi F^{*}_{mn}F^{mn} {+}
\left(\frac{1}{2} \eta_3 {-} \eta_2 \right) \nabla_m \phi \nabla^m
\phi \,, \label{Eineq1}
\end{equation}
with the tensors $T^{({\rm EM})}_{ik}$ and $T^{({\rm A})}_{ik}$
given by (\ref{TEM}) and (\ref{TAX}), respectively. The
non-minimal extension of the stress-energy tensor contains two
contributions: first, $T^{({\rm NMEM})}_{ik}$ describing pure
non-minimal electromagnetic part (see, e.g., \cite{BL05} for
details); second, the non-minimal axion part ${\cal T}^{({\rm
NMA})}_{ik}$. These tensors can be specified as follows:
\begin{equation}
T^{({\rm NMEM})}_{ik} = q_1 T^{(1)}_{ik} + q_2 T^{(2)}_{ik}+ q_3
T^{(3)}_{ik} \,, \label{decompEM}
\end{equation}
\begin{equation}
{\cal T}^{({\rm NMA})}_{ik} = Q_1 {\cal T}^{(1)}_{ik} + Q_2 {\cal
T}^{(2)}_{ik}+ Q_3 {\cal T}^{(3)}_{ik} + \eta_1 {\cal
T}^{(4)}_{ik} + \eta_2 {\cal T}^{(5)}_{ik} + \eta_3 {\cal
T}^{(6)}_{ik} + \eta_{({\rm A})} {\cal T}^{(7)}_{ik} \,,
\label{decompAX}
\end{equation}
where
\begin{equation} T^{(1)}_{ik} = - R F_{im}F_{k}^{ \ m}   +
\frac{1}{2} \left[ \nabla_{i} \nabla_{k} - g_{ik} \nabla^l
\nabla_l \right] \left[F_{mn}F^{mn} \right] \,, \label{T1}
\end{equation}
$$
T^{(2)}_{ik} = -\frac{1}{2}g_{ik} \left[\nabla_{m}
\nabla_{l}\left(F^{mn}F^{l}_{\ n} \right) - R_{lm} F^{mn}
F^{l}_{\ n} \right]
$$
$$
- F^{ln} \left(R_{il}F_{kn} + R_{kl}F_{in} \right) - R^{mn}
F_{im} F_{kn} - \frac{1}{2} \nabla^m \nabla_m \left(F_{in}
F_{k}^{ \ n}\right)
$$
\begin{equation}
\quad{}+\frac{1}{2}\nabla_l \left[ \nabla_i \left( F_{kn}F^{ln}
\right) + \nabla_k \left(F_{in}F^{ln} \right) \right] \,,
\label{T2}
\end{equation}%
$$
T^{(3)}_{ik} = \frac{1}{4}g_{ik} R^{mnls}F_{mn}F_{ls}-
\frac{3}{4} F^{ls} \left(F_{i}^{\ n} R_{knls} + F_{k}^{\
n}R_{inls} \right)
$$
\begin{equation}
\quad {}-\frac{1}{2} \nabla_{m} \nabla_{n} \left[ F_{i}^{ \
n}F_{k}^{ \ m} + F_{k}^{ \ n} F_{i}^{ \ m} \right] \,, \label{T3}
\end{equation}
\begin{equation}
{\cal T}^{(1)}_{ik}  \equiv  \frac{1}{2} \left[ \nabla_{i}
\nabla_{k} {-} g_{ik} \nabla^l \nabla_l \right] \left[\phi
F^{*}_{mn}F^{mn} \right] {-} \frac{1}{4} g_{ik} R \phi
F^{*}_{mn}F^{mn}\,, \label{calT1}
\end{equation}
$$
{\cal T}^{(2)}_{ik} \equiv {-} \frac{1}{2} \phi F^{*mn}
\left(R_{im}F_{kn} {+} R_{km}F_{in} \right) {+}
$$
$$
{+}\frac{1}{4}\nabla_l \left\{ \nabla_i \left[ \phi \left(
F_{kn}F^{*ln} {+} F^{*}_{kn}F^{ln} \right) \right] {+} \nabla_k
\left[ \phi \left( F_{in}F^{*ln} {+} F^{*}_{in}F^{ln} \right)
\right] \right\} {-}
$$
\begin{equation}
{-}\frac{1}{4}g_{ik} \nabla_{n} \nabla_{m}\left[\phi \left(
F^{nl}F^{*m}_{\ \ \ l} {+} F^{*nl}F^{m}_{\ \  l} \right)\right]
{-} \frac{1}{4} \nabla^m \nabla_m \left[ \phi \left(F_{in}
F_{k}^{ * n} {+} F^{*}_{in} F_{k}^{\ n} \right)\right]
 \,, \label{calT2}
\end{equation}
\begin{equation}
{\cal T}^{(3)}_{ik}  \equiv {-}\frac{1}{2} \nabla_{m} \nabla_{n}
\left[ \phi \left(F_{\  \ \ i}^{* n} F_{\ \ k}^{m} {+} F_{\ \ \
k}^{* n} F_{\ \ i}^{m} \right)\right] {+} \frac{1}{4} \phi F^{*mn}
\left(F_{il} R^{l}_{ \ kmn} {+} F_{kl} R^{l}_{\ imn} \right)
 \,,
\label{calT3}
\end{equation}
$$
{\cal T}^{(4)}_{ik} \equiv  \frac{1}{2}g_{ik} \left(R^l_n {-}
\nabla^l \nabla_n \right) \left( F^{nm} \nabla_m \phi \nabla_l
\phi \right) {+} \frac{1}{2} R^l_n \nabla_l \phi  \left(F_i^{\ n}
\nabla_k \phi  {+} F_k^{\ n} \nabla_i \phi    \right) {+}
$$
$$
{+}\frac{1}{4}\nabla^l \nabla_l \left[ \nabla_m \phi \left(
 F^m_{\ \ i} \nabla_k \phi {+}  F^m_{\ \ k} \nabla_i \phi
\right)\right] {+} \frac{1}{4}\nabla^l \left[ \nabla_i
\left(F_k^{\ m}\nabla_m \phi \nabla_l \phi \right) {+} \nabla_k
\left(F_i^{\ m}\nabla_m \phi \nabla_l \phi \right) \right] {+}
$$
$$
{+}\frac{1}{4}\nabla_m \left[ \nabla_i \left( F^{mn}\nabla_k \phi
\nabla_n \phi  \right) {+} \nabla_k \left( F^{mn}\nabla_i \phi
\nabla_n \phi  \right) \right] {+}
$$
\begin{equation}
{+}\frac{1}{2} F^{mn}   \left(R_{in}\nabla_k \phi {+}
R_{kn}\nabla_i \phi \right)\nabla_m \phi {+}
\frac{1}{2}\left(R^m_i F^n_{\ k} {+} R^m_k F^n_{\ i} \right)
\nabla_m \phi \nabla_n \phi \,, \label{calT4}
\end{equation}
\begin{equation}\label{calT5}
{\cal T}^{(5)}_{ik}=  R \nabla_i \phi \nabla_k \phi + \left(g_{ik}
\nabla_n \nabla^n - \nabla_i \nabla_k \right)\left[\nabla_m \phi
\nabla^m \phi \right]\,,
\end{equation}
$$
{\cal T}^{(6)}_{ik} = \nabla_m \phi \left[R_i^m \nabla_k \phi +
R_k^m \nabla_i \phi \right] - \frac{1}{4} R g_{ik} \nabla_m \phi
\nabla^m \phi +
$$
\begin{equation}\label{calT6}
+ \frac{1}{2}g_{ik} \nabla_m \nabla_n \left[ \nabla^m \phi
\nabla^n \phi \right] - \nabla^m \left[ \nabla_m \phi
\left(\nabla_i \nabla_k \phi \right) \right] \,.
\end{equation}
\begin{equation}\label{calT7}
{\cal T}^{(7)}_{ik}= \left(\nabla_i \nabla_k - g_{ik} \nabla_m
\nabla^m \right){\phi}^2\,.
\end{equation}
Straightforward calculations show that the following identity
takes place:
\begin{equation}
\nabla^k \left\{ \left(1{+}\kappa \Theta \right)^{-1} \left[
T^{({\rm EM})}_{ik}  {+} T^{({\rm A})}_{ik} {+} T^{({\rm
NMEM})}_{ik} {+} {\cal T}^{({\rm NMA})}_{ik} \right] \right\} = 0
\,, \label{EinBI}
\end{equation}
i.e. the total effective stress-energy tensor written in the
braces is a conserved quantity.

\subsection{Non-minimal constitutive equations for
electromagnetic field}

Relation (\ref{inducnm}) is a linear constitutive equation
\cite{Mauginbook,HehlObukhov,8Ni} of the following type:
\begin{equation}
H^{ik} = {\cal C}^{ikmn} F_{mn} \label{link}
\end{equation}
where
\begin{equation}
{\cal C}^{ikmn} = g^{ikmn} + \frac{1}{2} \phi \epsilon^{ikmn} +
{\cal R}^{ikmn} + \frac{1}{2} \phi \left[ \chi^{*ikmn}_{({\rm A})}
+ ^{*}\chi^{ikmn}_{({\rm A})}\right] \,. \label{nminduc}
\end{equation}
Here we use the standard definitions for the right and left
dualization:
\begin{equation}
\chi^{*ikmn}_{({\rm A})} \equiv \frac{1}{2}\chi^{ikpq}_{({\rm
A})} \epsilon_{pq}^{\ \ mn} \,, \quad ^{*}\chi^{ikmn}_{({\rm A})}
\equiv \frac{1}{2} \epsilon^{ik}_{\ \ pq} \chi^{pqmn}_{({\rm A})}
\,. \label{nminduc12}
\end{equation}
The tensor ${\cal C}^{ikmn}$ describes the linear response of the
material to the electromagnetic field action and contains the
information about non-minimal dielectric permittivity and magnetic
impermeability, as well as about the non-minimal magneto-electric
coefficients in analogy with the standard continuum
electrodynamics \cite{Mauginbook,landau,nu1}.

Let us mention that in many works (e.g., in \cite{HehlObukhov},
\cite{nobi1}-\cite{nobi3}), a constitutive tensor density
$\hat{\chi}^{ikmn} = \sqrt{-g} \ {\cal C}^{ikmn}$ is used for the
description of linear response instead of true tensor ${\cal
C}^{ikmn}$. In terms of the quantity $\hat{\chi}^{ikmn}$ the
condition for no birefringence (no splitting, no retardation) for
electromagnetic wave propagation in all directions in the weak
field limit gives ten constraint equations on the components of
constitutive tensor density (see, e.g.,
\cite{nobi1}-\cite{nobi4}). With these ten constraints
$\hat{\chi}^{ikmn}$ can be rewritten in the following form
\begin{equation}
\hat{\chi}^{ikmn} = \frac{1}{2}\sqrt{-H} \ \psi
\left[H^{im}H^{kn}-H^{in}H^{km} \right] + \varphi E^{ikmn}  \,,
\label{nobiref}
\end{equation}
where $H={\rm det}(H_{ik})$ is a determinant of an effective
metric $H_{ik}$, which generates the light cone for
electromagnetic wave propagation; $\psi$ is some dilation factor;
$\varphi$ differs from our $\phi$ by the coefficient two $\phi = 2
\varphi$. In case when (\ref{nobiref}) is not satisfied,
birefringence will occur. In case when (\ref{nobiref}) is
satisfied, a pseudoscalar field $\varphi$ will give a polarization
rotation (an optical activity induced by $\varphi$). The effects
of birefringence and polarization rotation described in terms of
constitutive tensor density are studied in the 1970's and 1980's.
In particular, constraints on birefringence from pulsar signal
observations give the following estimates for the birefringence
part $\delta \hat{\chi}^{ikmn}$ of  $\hat{\chi}^{ikmn}$ in a weak
field:
\begin{equation}
\left|\delta \hat{\chi}^{ikmn}\right| < 10^{-14} - 10^{-16}\,.
\label{nobiref1}
\end{equation}
Using the medium velocity four-vector $U^i$, normalized by
$U^iU_i=1$, one can decompose ${\cal C}^{ikmn}$ uniquely as
\begin{eqnarray}
&{\cal C}^{ikmn} = \frac12 \left( \varepsilon^{im} U^k U^n -
\varepsilon^{in} U^k U^m + \varepsilon^{kn} U^i U^m -
\varepsilon^{km} U^i U^n \right) + \nonumber \\& +\frac12 \left[
-\eta^{ikl}(\mu^{-1})_{ls}  \eta^{mns} + \eta^{ikl}(U^m\nu_{l}^{\
n} - U^n \nu_{l}^{\ m}) + \eta^{lmn}(U^i \nu_{l}^{\ k} - U^k
\nu_{l}^{\ i} ) \right] \,. & \label{44}
\end{eqnarray}
Here $\varepsilon^{im}$ is the dielectric permittivity tensor,
$(\mu^{-1})_{pq}$ is the magnetic impermeability tensor, and
$\nu_{p \ \cdot}^{\ m}$ is the tensor of magneto-electric
coefficients. These quantities are defined as follows:
$$
\varepsilon^{im} = 2 {\cal C}^{ikmn} U_k U_n \,, \quad
(\mu^{-1})_{pq}  = - \frac{1}{2} \eta_{pik} {\cal C}^{ikmn}
\eta_{mnq}\,,
$$
\begin{equation}
\nu_{p}^{\ m} = \eta_{pik} {\cal C}^{ikmn} U_n =U_k {\cal
C}^{mkln} \eta_{lnp}\,. \label{varco}
\end{equation}
We use the symbols $\eta_{mnl}$ and $\Delta^{ik}$ for the tensors
\begin{equation}
\eta_{mnl} \equiv \epsilon_{mnls} U^s \,,
\quad
\Delta^{ik} = g^{ik} - U^i U^k \,,
\label{47}
\end{equation}
orthogonal to $U^i$.
The tensors $\varepsilon_{ik}$ and $(\mu^{-1})_{ik}$ are
symmetric, but $\nu_{lk}$ is in general non-symmetric. These three
tensors are orthogonal to $U^i$,
\begin{equation}
\varepsilon_{ik} U^k = 0, \quad (\mu^{-1})_{ik} U^k = 0, \quad
\nu_{l}^{\ k} U^l = 0 = \nu_{l}^{\ k} U_k \,. \label{orthog}
\end{equation}
Using expression (\ref{nminduc}) one can calculate the tensors
$\varepsilon^{im}$, $(\mu^{-1})_{im}$ and $\nu^{pm}$ explicitly.
The symmetric dielectric permittivity tensor
$$
\varepsilon^{im} = \Delta^{im}\left[ 1{+}
q_1R {+} q_2 R^{pq}U_pU_q \right] + q_2
R_{pq}\Delta^{ip}\Delta^{mq} {+}
$$
\begin{equation}
{+} 2q_3 R^{ipmq}U_pU_q {+}Q_3  \phi \ U_p U_q \left(^{*} R^{ipmq}
{+} R^{*ipmq} \right) \,, \label{eps}
\end{equation}
as well as the symmetric magnetic impermeability tensor
$$
\left(\mu^{-1}\right)^{im} = \Delta^{im}\left[ 1{+} q_1R {+} q_2
R^{pq}U_pU_q \right] {-} q_2 R_{pq}\Delta^{ip}\Delta^{mq} {-}
$$
\begin{equation}
{-} 2q_3 \  ^{*}R^{*ipmq}U_pU_q {+}Q_3  \phi \ U_p U_q \left(^{*}
R^{ipmq} {+} R^{*ipmq} \right)  \label{muu}
\end{equation}
do not contain coupling parameters $Q_1$ and $Q_2$. The
cross-tensor
$$
\nu^{pm} =  q_2 \ \eta^{pmk}R_{kl}U^l + 2q_3  \ ^{*}R^{plmn}U_l
U_n +
$$
\begin{equation}
 - \Delta^{pm}\phi \left[1{+} \left(Q_1 {+}
\frac{1}{2}Q_2 \right) R \right] {+}Q_3 \phi U_l U_n
\left(^{*}R^{*plmn} {-} R^{plmn} \right) \,, \label{nuu}
\end{equation}
which gives the optical activity effects (see, e.g., \cite{nu1}),
contains two explicitly distinguished parts. The first part does
not contain pseudoscalar field $\phi$, is linear in the parameters
$q_2$ and $q_3$ and describes optical activity induced by direct
non-minimal interaction between electromagnetic and gravitational
fields (such a model was analyzed in \cite{BL2,BDZ2}). It is
important to mention that this part of the cross-tensor $\nu^{pm}$
contains both: skew-symmetric and symmetric terms (see the terms
with $\eta_{pmk}$ and $^{*}R^{plmn}U_l U_n$). The second
contribution to $\nu^{pm}$, which is proportional to the
pseudoscalar field $\phi$, is symmetric with respect to
transposition of indices (see the terms, containing the projector
$\Delta^{pm}$ and the tensor $^{*}R^{*plmn} {-} R^{plmn}$). It
describes an optical activity induced by the axion field, the term
$\Delta^{pm} \phi$ relates to the minimal effects, the terms
proportional to $Q_1$, $Q_2$ and $Q_3$ relates to non-minimal
effects. It is well-known that the non-vanishing cross-tensor
$\nu^{ik}$ indicates that the medium is optically active, and the
rotation of the Faraday type takes place in the course of
electromagnetic wave propagation. Thus, one can see directly from
(\ref{nuu}), that the non-minimal interaction between
electromagnetic and axion fields produces curvature-induced
optical activity of a new type. Below we consider this effect
explicitly by the example of model with pp-wave symmetry.

\noindent In table 1, we summarize briefly the information
about non-minimal coupling constants.

\vspace{3mm}

\begin{tabular}{|c|l|l|}
  \hline
     &  Term in the  & Physical meaning \\
    & Lagrangian & \\
\hline \hline
   $q_1$ & $\frac{1}{2} R F^{mn} F_{mn}$ & {\small NMEM susceptibility linear in the Ricci scalar} \\
  \hline
    $q_2$ & $ R^{mn} F_{mk} F_{n}^{\ k}$ & {\small NMEM susceptibility linear in the Ricci tensor} \\
    \hline
    $q_3 $ &  $ \frac{1}{2} R^{ikmn} F_{ik} F_{mn}$ & {\small NMEM susceptibility linear
    in the Riemann tensor} \\
\hline \hline
$Q_1$ & $\frac{1}{2} \phi R F^{mn} F^{*}_{mn}$ & {\small NMEM susceptibility induced by the axion with the Ricci scalar}  \\
  \hline
    $Q_2$ & $ \phi R^{mn} F_{mk} F_{n}^{* k}$ & {\small NMEM susceptibility induced by the axion with the Ricci tensor} \\
    \hline
    $Q_3 $ &  $ \frac{1}{2} \phi R^{ikmn} F_{ik} F^{*}_{mn}$ & {\small NMEM susceptibility induced by the axion with the Riemann tensor} \\
\hline \hline $\eta_1$ & ${-}   R^n_{\ l} F^{ml}  \nabla_m \phi
\nabla_n
\phi $ & {\small NMEM current induced by the axion field gradient} \\
  \hline
    $\eta_2$ & ${-} R \nabla^m \phi  \nabla_m \phi$ & {\small NM axion-graviton derivative coupling with the Ricci scalar} \\
    \hline
    $\eta_3 $ &  ${-} R^{mn} \nabla_m \phi  \nabla_n \phi$ & {\small NM axion-graviton derivative coupling with the Ricci tensor}  \\
\hline \hline
$\eta_{({\rm A})}$ & $R \phi^2$ & {\small NM correction to the mass square of the axion} \\
\hline
\end{tabular}

\vspace{2mm} \noindent {\small Table I. Ten non-minimal (NM)
coupling parameters are divided into four subgroups: the first
$q_1,q_2,q_3$; second $Q_1,Q_2,Q_3$; third $\eta_1,\eta_2,
\eta_3$; and fourth $\eta_{({\rm A})}$. In the second column the
terms in the Lagrangian are given in front of which the
corresponding coupling parameters are introduced; the parameters
of the first subgroup introduce the terms without the pseudoscalar
field $\phi$; the parameters of the second subgroup relate to the
terms linear in $\phi$; the terms indicated by $\eta_1,\eta_2,
\eta_3$, are quadratic in the four-gradient of $\phi$; and
finally, $\eta_{({\rm A})}$ introduces the term quadratic in
$\phi$. In the last column, we point out the physical meaning of
these non-minimal terms, based on decompositions of the
constitutive tensors for the electromagnetic (EM) and pseudoscalar
fields. }


\section{Non-minimal models with pp-wave symmetry}

\subsection{Reduced non-minimal field equations}

We consider the line element for the spacetime with pp-wave
symmetry in the form \cite{BPR}
\begin{equation}
\mbox{d}s^{2} = 2\mbox{d}u\mbox{d}v - L^{2} \left\{\cosh2\gamma
\left[e^{2 \beta}(\mbox{d}x^2)^2 + e^{-2\beta} (\mbox{d}x^3)^2
\right]+ 2 \sinh{2\gamma} \mbox{d}x^2 dx^3 \right\} \,,
\label{ppmetric}
\end{equation}
where $u {=} \frac{ct {-} x^1}{\sqrt{2}}$ and $v {=} \frac{ct {+}
x^1}{\sqrt{2}}$ are the retarded and the advanced times,
respectively; $\beta(u)$ and $\gamma(u)$ are functions of retarded
time only, they determine the polarization of gravitational wave
field. The case when the gravitational pp-wave field is
characterized by one polarization only ($\gamma=0$) is
investigated in detail in \cite{MTW}. Various properties of
pp-waves are discussed, e.g., in \cite{Petrov,Synge,Ex}. For the
metric (\ref{ppmetric}), the Riemann tensor has only three
non-vanishing components $R_{2u2u}$, $R_{3u3u}$ and $R_{2u3u}$.
The Ricci tensor has only one (generally) non-vanishing component
$R_{uu}$ and the Ricci scalar is equal to zero identically $R=0$.
For this metric the three Killing vectors, which form an Abelian
subgroup of the group $G_5$ \cite{Ex}, are
\begin{equation}
\xi_{(v)}^i = \delta_v^i, \quad \xi_{(2)}^i = \delta_2^i,\quad
\xi_{(3)}^i = \delta_3^i \ . \label{Kill}
\end{equation}
The Killing vector $\xi_{(v)}^i$ is a covariant constant null
vector orthogonal to $\xi_{(2)}^i$ and $\xi_{(3)}^i$. We suggest
here that the axion and electromagnetic field
potentials inherit the spacetime symmetry, i.e. the Lie
derivatives satisfy the following relations:
\begin{equation}
\pounds_{\xi_{(a)}} \phi = 0 \,, \quad \pounds_{\xi_{(a)}}{\cal
A}_{m} =0 \,, \label{Lie}
\end{equation}
where $(a)$ take the values $(v)$, $(2)$ and $(3)$. As a
consequence of (\ref{Lie}), the pseudoscalar field and
electromagnetic field potential depend on the retarded time only,
$\phi = \phi(u)$, ${\cal A}_{m} = {\cal A}_{m}(u)$. Taking into
account the Lorentz gauge $\nabla_k {\cal A}^k =0$, we can choose
the potential in the form
\begin{equation}
{\cal A}_{u} = {\cal A}_{v} = 0 \,, \quad {\cal A}_2= {\cal
A}_2(u)\,, \quad {\cal A}_3={\cal A}_3(u) \,. \label{Auv23}
\end{equation}
The next step is to verify that pp-wave symmetry is admitted by the
total system of non-minimal master equations. Direct calculations
show that for such an  ansatz about symmetry the non-minimal
equation for pseudoscalar field $\phi(u)$, i.e. the equations
(\ref{eqaxi1}) with (\ref{sus3}), are satisfied identically for
arbitrary $Q_1$,...$\eta_3$ and $\eta_{({\rm A})}$, when the axion
field is massless, i.e. $m_{({\rm A })}=0$. Also, the
non-minimal electrodynamic equations (\ref{max2}) with
(\ref{inducnm}) and (\ref{current}) are satisfied identically for
arbitrary coupling constants, when the potential takes the form
(\ref{Auv23}). Finally, the non-minimal equations for the
gravitational field (\ref{Eineq}) with
(\ref{Eineq1})-(\ref{calT7}) can be reduced to one equation only,
namely
$$
{-}\frac{2}{\kappa} \left[\frac{L^{\prime \prime}}{L} {+}
\left(\beta^{\prime}\right)^2 \cosh^2{2\gamma} {+}
\left(\gamma^{\prime}\right)^2 \right] \left[1 {+} \kappa
\eta_{({\rm A})} \phi^2 \right] {=} \left(\phi^{\prime}\right)^2
\left(1 {+} 2\eta_{({\rm A})} \right) {+} 2\eta_{({\rm A})} \phi
\phi^{\prime \prime} {+}
$$
\begin{equation}
{+}\frac{1}{L^2} \left\{ \cosh{2\gamma} \left[\left({\cal
A}^{\prime}_2 e^{{-}\beta}\right)^2 {+} \left({\cal A}^{\prime}_3
e^{\beta}\right)^2 \right] {-} 2\sinh{2\gamma} \ {\cal
A}^{\prime}_2 {\cal A}^{\prime}_3 \right\} \,. \label{Lpp}
\end{equation}
The prime denotes here and below the derivative with respect to
the retarded time $u$. The right-hand side of this equation
includes, as usual, the minimal contributions from the tensors
$T^{({\rm A})}_{ik}$ and $T^{({\rm EM})}_{ik}$. Only two
non-minimal contributions came from the term ${\cal T}^{(7)}_{ik}$
(\ref{calT7}) and from the term $\kappa \eta_{({\rm A})} \phi^2$
in the expression for $\Theta$ (\ref{Eineq1}). The non-minimal
coupling constants $q_1$, $q_2$, $q_3$, $Q_1$ $Q_2$, $Q_3$,
$\eta_1$, $\eta_2$, $\eta_3$ are non-vanishing, but they happen to
be hidden in this pp-wave model. Thus, in the presented model, six
functions $\phi(u)$, ${\cal A}_2(u)$, ${\cal A}_3(u)$, $\beta(u)$,
$\gamma(u)$ and $L(u)$ are linked by one equation (\ref{Lpp})
only, and this model admits very wide possibilities in searching
for exact solutions.

\subsection{Exact solutions of the pp-wave type: \\
quasi-minimal models with $\eta_{({\rm A})}=0$}

When the non-minimal parameter $\eta_{({\rm A})}$ is vanishing, we
deal with a model which looks like minimal, i.e. the key
equation for the pp-wave gravitational field
$$
{-} \frac{L^{\prime \prime}}{L} = \left(\beta^{\prime}\right)^2
\cosh^2{2\gamma} {+} \left(\gamma^{\prime}\right)^2 {+}
\frac{\kappa}{2}\left(\phi^{\prime}\right)^2 {+}
$$
\begin{equation}
{+}\frac{\kappa}{2L^2} \left\{ \cosh{2\gamma} \left[\left({\cal
A}^{\prime}_2 e^{{-}\beta}\right)^2 {+} \left({\cal A}^{\prime}_3
e^{\beta}\right)^2 \right] {-} 2\sinh{2\gamma} \ {\cal
A}^{\prime}_2 {\cal A}^{\prime}_3 \right\}  \label{Lppmin}
\end{equation}
does not contain non-minimal parameters at all. That is why we
indicate this case with arbitrary but hidden parameters $q_1$,
$q_2$, $q_3$, $Q_1$ $Q_2$, $Q_3$, $\eta_1$, $\eta_2$, $\eta_3$ and
vanishing $\eta_{({\rm A})}$ as {\it quasi-minimal}. Clearly, the
quantity $\left(-\frac{L^{\prime \prime}}{L}\right)$ in the
left-hand side of (\ref{Lppmin}) has to be non-negative. This
property of $L(u)$, which has initial value $L(0)=1$, provides
$L(u)$ to vanish at some moment $u=u^*$ ($L(u^*)=0$)(see, e.g.,
\cite{Synge}). This moment of the retarded time relates to the
well-known singularity, the physical sense of which is discussed
in detail in many books. In \cite{Ex}, a number of papers are
quoted, in which exact solutions of (\ref{Lppmin}) are obtained
for the pure electromagnetic source ($\phi =0$). In addition,
exact solutions of the model of this type are known for vanishing
${\cal A}_k$, when the (pure)scalar field $\Phi$ is the source of
the pp-wave gravity field (see, e.g., \cite{Aon1}). Finally, when
${\cal A}_k=0$ and $\phi =0$, we deal with the so-called pure
gravitational pp-wave (see, e.g., \cite{MTW,Petrov,Synge,Ex}). We
would like to list here only three examples from the wide
collection of minimal solutions of the equation (\ref{Lppmin}),
which relates to the models with symmetric spacetime of the
pp-wave type. The models with symmetric spacetime are
characterized by the condition (\cite{Petrov,Ex})
\begin{equation}
\nabla_k R^i_{\ mnl} = 0 \,. \label{qq13}
\end{equation}
In particular, when $\gamma(u)=0 $, one obtains from (\ref{qq13})
that
\begin{equation}
R^2_{\ u2u} = - \left[\frac{L^{\prime \prime}}{L} +
(\beta^{\prime})^2 \right] - \left[2
\frac{L^{\prime}}{L}\beta^{\prime} + \beta^{\prime \prime} \right]
= \lambda_1 \,, \label{qq14}
\end{equation}
\begin{equation}
R^3_{\ u3u} = - \left[\frac{L^{\prime \prime}}{L} +
(\beta^{\prime})^2 \right] + \left[2
\frac{L^{\prime}}{L}\beta^{\prime} + \beta^{\prime \prime} \right]
= \lambda_2 \,, \label{qq15}
\end{equation}
where $\lambda_1$ and $\lambda_2$ are some constants. The
summation of (\ref{qq14}) and (\ref{qq15}) yields
\begin{equation}
R_{uu} = - \left[\frac{L^{\prime \prime}}{L} + (\beta^{\prime})^2
\right] = \lambda_1 + \lambda_2 \,. \label{qq16}
\end{equation}
Thus, for the symmetric spacetime, the key equation for the
(quasi)minimal model can be reduced to
\begin{equation}
\frac{2}{\kappa} \left( \lambda_1 {+} \lambda_2 \right)  {=}
\left(\phi^{\prime}\right)^2 {+} \frac{1}{L^2}\left[\left({\cal
A}^{\prime}_2 e^{{-}\beta}\right)^2 {+} \left({\cal A}^{\prime}_3
e^{\beta}\right)^2 \right]  \,. \label{e34qq}
\end{equation}
In more appropriate terms, when
\begin{equation}
L^2 = F(u) \ G(u) \,, \quad 2\beta = \log{\frac{F}{G}} \,,
\label{qsq1}
\end{equation}
(\ref{qq14}), (\ref{qq15}) give two independent equations
\begin{equation}
F^{\prime \prime}(u) + \lambda_1 F = 0 \,, \quad G^{\prime
\prime}(u) + \lambda_2 G = 0 \,, \label{qsq2}
\end{equation}
which can be easily solved if the signs of the constants
$\lambda_1$ and $\lambda_2$ are fixed.

\vspace{3mm}

\noindent (i) {\it First example}

\noindent The first model of such type relates to the pure
gravitational wave, i.e. to the case when $R_{uu}=0$. Suggesting
that $\lambda_1=-\lambda_2 \equiv \nu^2$, we obtain immediately
from (\ref{qsq2}) the well-known Petrov's solution \cite{Petrov}
\begin{equation}
\mbox{d}s^{2} = 2\mbox{d}u\mbox{d}v {-} \cos^2{\nu u}
(\mbox{d}x^2)^2 {-} \cosh^2{\nu u} (\mbox{d}x^3)^2  \,,
\label{330}
\end{equation}
with
\begin{equation}
F(u) = \cos{\nu u} \,, \quad G(u) = \cosh{\nu u}  \,. \label{331}
\end{equation}
This solution is admissible, when $\phi {=} \phi_0 {=} const$,
${\cal A}_k {=}0$. The quantity $\sqrt{{-}g} \equiv \cos{\nu u}
\cosh{\nu u}$ vanishes at $u^* = \pi / 2 \nu$.

\vspace{3mm}

\noindent (ii) {\it  Second example}

\noindent When  ${\cal A}_2 = {\cal A}_3 = 0$, $\beta = \gamma =
0$ and function $\phi(u)$ is linear in the retarded time, i.e.
$\phi(u) = \phi_0 + \omega u$, then (\ref{Lppmin}) gives
\begin{equation}
L(u) = \cos{\sqrt{\frac{\kappa}{2}}\omega u} \,, \quad L(0) = 1
\,, \quad L^{\prime}(0)=0 \,. \label{nm12}
\end{equation}
For such a metric one obtains that
\begin{equation}
R^2_{\ u2u} = R^3_{\ u3u} = - \frac{L^{\prime \prime}}{L} =
\frac{\kappa \omega^2}{2} = \lambda_1 = \lambda_2 \,. \label{nm13}
\end{equation}
The first zero of the function $L$ is $u=u^*=\frac{\pi}{\omega
\sqrt{2\kappa}}$.

\vspace{3mm}

\noindent (iii) {\it Third example}

\noindent When $\lambda_1=\nu^2$, $\lambda_1=- \mu^2$ and
$\phi=0$, there is an exact solution
\begin{equation}
F(u)= \cos{\nu u} \,, \quad G(u) = \cosh{\mu u}  \,, \label{nm14}
\end{equation}
with
\begin{equation}
\nu^2 = \mu^2 + \frac{\kappa E_0^2}{2}\,. \label{nm15}
\end{equation}
The electromagnetic field in this model is presented by the following
potentials:
\begin{equation}
{\cal A}_2(u) {=} \frac{E_0}{(\omega^2 {-}
\nu^2)}\left[ \omega \sin{\omega u} \cos{\nu u}  {-} \nu \sin{\nu
u} \cos{\omega u} \right]\,, \label{eqq12}
\end{equation}
\begin{equation}
{\cal A}_3(u) {=} \frac{E_0}{(\omega^2 {+} \mu^2)}\left[ \mu
\sin{\omega u} \sinh{\mu u}  {-} \omega \cos{\omega u} \cosh{\mu
u} \right] \,. \label{eqq13}
\end{equation}
This electromagnetic wave can be considered as a circularly
polarized wave, since the physical components of the electric
field satisfy the following conditions:
\begin{equation}
E^2_{\bot} \equiv -(E_2E^2 + E_3E^3) \equiv - g^{\alpha \beta}
{\cal A}^{\prime}_{\alpha} {\cal A}^{\prime}_{\beta} = E^2_0 \,,
\quad (\alpha, \beta = 2,3) \,. \label{circ}
\end{equation}
When $\eta_{({\rm A})}\neq 0$, the quantity $\left(-\frac{L^{\prime
\prime}}{L}\right)$ can change the sign, since the term
$\eta_{({\rm A})} (\phi^2)^{\prime \prime}$ in the right-hand side
of (\ref{Lpp}) is not positively defined in general case. This
means that the singularity $L=0$ can, in principle, be avoided.

\subsection{Exact solutions of the pp-wave type: \\ Non-minimal models with $\eta_{({\rm A})} \neq 0$}

Consider now exactly solvable models with non-vanishing coupling
constant $\eta_{({\rm A})}$.

\subsubsection{Regular model}

Let us consider a  model with $L(u) \equiv 1$ and $\gamma(u)
\equiv 0 $. It can be indicated as the regular one, since
$det(g_{ik})= -L^4 \equiv -1$ and can not vanish. Equation for
$\beta$ reduces to
\begin{equation}
{-}\frac{2}{\kappa}  \left(\beta^{\prime}\right)^2 \left[1 {+}
\kappa \eta_{({\rm A})} \phi^2 \right] {=}
\left(\phi^{\prime}\right)^2 {+} \eta_{({\rm A})}
\left(\phi^2\right)^{\prime \prime} {+}
 \left({\cal A}^{\prime}_2 e^{{-}\beta}\right)^2 {+} \left({\cal
A}^{\prime}_3 e^{\beta}\right)^2   \,. \label{e358}
\end{equation}
When $\eta_{({\rm A})}=0$ there is no real solutions of this
equation, but such a possibility appears in the non-minimal case.
We consider only one example of the exact regular models, it is
characterized by
\begin{equation}
 \phi = \phi_0  \,,  \quad {\cal A}_2(u)= {\cal
A}_2(0) e^{\beta(u)} \,, \quad {\cal A}_3(u)= {\cal A}_3(0) e^{-
\beta(u)} \,, \label{e0358}
\end{equation}
and is possible, when $\eta_{({\rm A})}<0 $ and
\begin{equation}
\phi^2_0 \ |\eta_{({\rm A})}| = 1 + \frac{\kappa}{2}\left[ {\cal
A}^2_2(0) + {\cal A}^2_3(0)\right] \,. \label{e243}
\end{equation}
The function $\beta(u)$ is arbitrary; we prefer to use the
following periodic finite function:
\begin{equation}
\beta(u) = \frac{1}{2} \beta_{({\rm max })} (1-\cos{2\lambda u})
\,, \quad \beta(0) = 0 \,, \quad \beta^{\prime}(0) = 0 \,.
\label{e343}
\end{equation}
The metric for this non-minimal model is regular:
\begin{equation}
\mbox{d}s^{2} {=} 2\mbox{d}u\mbox{d}v {-} \left\{\exp
\left[2\beta_{({\rm max })}\sin^2{\lambda u}\right]
(\mbox{d}x^2)^2 {+} \exp \left[{-}2\beta_{({\rm max
})}\sin^2{\lambda u}\right] (\mbox{d}x^3)^2 \right\} \,,
\label{30}
\end{equation}
the potentials of the electromagnetic field and their derivatives
are also periodic and regular.

\subsubsection{Non-minimal models with symmetric spacetime of the pp-wave type}

The key equation for the metric coefficients can be reduced in
this case to the relation
\begin{equation}
\frac{2}{\kappa} \left( \lambda_1 {+} \lambda_2 \right) \left[1
{+} \kappa \eta_{({\rm A})} \phi^2 \right] {=}
\left(\phi^{\prime}\right)^2 \left(1 {+} 2\eta_{({\rm A})}\right)
{+} 2\eta_{({\rm A})} \phi \phi^{\prime \prime} {+}
\frac{1}{L^2}\left[\left({\cal A}^{\prime}_2 e^{{-}\beta}\right)^2
{+} \left({\cal A}^{\prime}_3 e^{\beta}\right)^2 \right]  \,.
\label{Ne34qq}
\end{equation}
Let us consider three exact solutions of this equation, which can
be indicated as the non-minimal ones.

\vspace{3mm}

\noindent (1) {\it First exact solution}

\noindent Let us suppose that $\lambda_1=-\lambda_2 \equiv
\nu^2$,  and we again deal with the Petrov solution (\ref{330}).
This solution is admissible when
\begin{equation}
\left\{\left(\phi^{\prime}\right)^2 \left(1 {+} 2 \eta_{({\rm
A})}\right) {+} 2\eta_{({\rm A})} \phi \phi^{\prime \prime}
\right\} \cos^2{\nu u} \cosh^2{\nu u} {+} \left[\left({\cal
A}^{\prime}_2 \cosh{\nu u}\right)^2 {+} \left({\cal A}^{\prime}_3
\cos^2{\nu u}\right)^2 \right] =0  \,. \label{e34qq1}
\end{equation}
One of the solutions of this equation, which relates to the case
$\eta_{({\rm A})} {=} {-} \frac{1}{4}$, is given by
\begin{equation}
\phi(u) {=} \phi_0 \cosh{\sigma u} \,,  \quad \phi_0 {=} \pm
\sqrt2 \ \frac{E_0}{\sigma} \,, \label{eqq1}
\end{equation}
\begin{equation}
{\cal A}_2(u) {=} \frac{E_0}{(\omega^2 {-} \nu^2)}\left[ \omega
\sin{\omega u} \cos{\nu u}  {-} \nu \sin{\nu u} \cos{\omega u}
\right]\,, \label{Reqq12}
\end{equation}
\begin{equation}
{\cal A}_3(u) {=} \frac{E_0}{(\omega^2 {+} \nu^2)}\left[ \nu
\sin{\omega u} \sinh{\nu u}  {-} \omega \cos{\omega u} \cosh{\nu
u} \right] \,. \label{Reqq13}
\end{equation}
Here $E_0$, $\sigma$, $\omega$ and $\nu$ are arbitrary constant.
When $\omega^2 = \nu^2$, the expression for ${\cal A}_2(u)$ should
be replaced by ${\cal A}_2(u) = \frac{E_0}{4\nu} (2 \nu u +
\sin{2\nu u})$. Let us mention that in the paper \cite{Aon1}, the
case $\xi = 1/4$ is also indicated as the special one
($\eta_{({\rm A})}$ in our notations corresponds to $\xi$ in
\cite{Aon1}).

\vspace{3mm}

\noindent (2) {\it  Second exact solution}

\noindent In the case $\lambda_1 = \lambda_2 = - \mu^2$, an
appropriate example of the model is characterized by
\begin{equation}
\beta(u) = 0 \,, \quad L(u) = F(u) = G(u) = \cosh{\mu u} \,, \quad
L(0)=1 \,, \quad L^{\prime}(0) =0 \,, \label{nqq14}
\end{equation}
\begin{equation}
R^2_{\ u2u} = R^3_{\ u3u} = - \frac{L^{\prime \prime}}{L} = -
\mu^2 \,, \label{nqq141}
\end{equation}
\begin{equation}
{\cal A}_2(u) {=} \frac{E_0}{(\omega^2 {+} \mu^2)}\left[ \omega
\sin{\omega u} \cosh{\mu u}  {+} \mu \sinh{\mu u} \cos{\omega u}
\right]\,, \label{neqq12}
\end{equation}
\begin{equation}
{\cal A}_3(u) {=} \frac{E_0}{(\omega^2 {+} \mu^2)}\left[ \mu
\sin{\omega u} \sinh{\mu u}  {-} \omega \cos{\omega u} \cosh{\nu
u} \right] \,, \label{neqq13}
\end{equation}
\begin{equation}
\phi(u) = \phi_0 = const \,, \quad \eta_{({\rm A})}  < 0 \,, \quad
\kappa E^2_0 = 4 \mu^2 \left( |\eta_{({\rm A})}| \kappa \phi^2_0 -
1 \right) > 0 \,. \label{ne34qq}
\end{equation}
Again, we deal with circularly polarized electromagnetic wave,
whose amplitude $E_0$ is connected with constant pseudoscalar
field by a special manner (\ref{ne34qq}), which is admissible at
the non-minimal case only.

\vspace{3mm}

\noindent (3) {\it  Third exact solution}

\noindent When $\lambda_1=\nu^2$, $\lambda_2=- \mu^2$, there is a
simple non-minimal modification of the solution (\ref{nm14}),
(\ref{nm15}) with constant non-vanishing pseudoscalar field
\begin{equation}
F(u)= \cos{\nu u} \,, \quad G(u) = \cosh{\mu u} \,, \quad
\phi(u)=\phi_0 \,, \label{Nnm14}
\end{equation}
\begin{equation}
\nu^2 = \mu^2 + \frac{\kappa E_0^2}{2 (1+ \kappa \eta_{(A)}
\phi_0^2)}\,, \label{Nnm15}
\end{equation}
with electromagnetic potentials, given by (\ref{eqq12}),
(\ref{eqq13}).

\subsubsection{Special model}

When a pseudoscalar field is constant, i.e. $\phi(u)=\phi_0$, and
electromagnetic field is absent, the equation for the
gravitational field takes the form
\begin{equation}
\left[\frac{L^{\prime \prime}}{L} {+}
\left(\beta^{\prime}\right)^2 \cosh^2{2\gamma} {+}
\left(\gamma^{\prime}\right)^2 \right] \left[1 {+} \kappa
\eta_{({\rm A})} \phi^2_0 \right] {=} 0
 \,.
\label{e349}
\end{equation}
When $\phi^2_0 = - 1 / \kappa \eta_{({\rm A})}$, the equations for
gravity field are satisfied for arbitrary $L(u)$, $\beta(u)$ and
$\gamma(u)$, and we deal with a non-trivial special case. It
corresponds to the case when the Ricci scalar disappears from the
Lagrangian.

\subsubsection{Cheshire smile}

When the spacetime is the Minkowski one, and $L=1$, $\beta=0$,
$\gamma=0$, the equation (\ref{Lppmin}) yields
\begin{equation}
0 {=} \left(\phi^{\prime}\right)^2 {+} \eta_{({\rm A})}
\left(\phi^2\right)^{\prime \prime} {+}
 \left({\cal
A}^{\prime}_2 \right)^2 {+} \left({\cal A}^{\prime}_3 \right)^2
\,, \label{e38}
\end{equation}
and this equation admits real solutions when the non-minimal
parameter $\eta_{({\rm A})}$ is non-vanishing. The exotic
situation when (pure) scalar field does not curve the spacetime,
is called in \cite{Che} as gravitational Cheshire smile. Here we
deal with pseudoscalar field instead of pure scalar one;
nevertheless, the physical sense of such a solution remains the
same as in \cite{Che}. When the electromagnetic field is described
by the circularly polarized wave with
\begin{equation}
{\cal A}_2(u) = \frac{E}{\lambda} \cos{\lambda u} \,, \quad {\cal
A}_3(u) = \frac{E}{\lambda} \sin{\lambda u} \,, \label{e389}
\end{equation}
then the pseudoscalar field satisfies the equation
\begin{equation}
\left(\phi^{\prime}\right)^2 {+} \eta_{({\rm A})}
\left(\phi^2\right)^{\prime \prime} {+} E^2  =0\,, \label{e41}
\end{equation}
whose solution can be represented in quadratures as
\begin{equation}
\pm (u+C_2) = \int\frac{d \phi}{\sqrt{C_1 \phi^{-
\frac{1+2\eta_{({\rm A})}}{\eta_{({\rm A})}}} -
\frac{E^2}{1+2\eta_{({\rm A})}} } } \,. \label{e42}
\end{equation}
One of the explicit particular solutions to (\ref{e42}) is
\begin{equation}
\phi = \phi_0 \cosh ku \,, \quad \eta_{({\rm A})} = - \frac{1}{4}
\,, \quad \phi_0 = \sqrt{2} \  \frac{E}{k} \,. \label{e43}
\end{equation}
Again we used the special value of the non-minimal parameter
$\eta_{({\rm A})} = - \frac{1}{4}$ as in \cite{Aon1}. This
solution is infinite at $u \to \infty$.


\noindent In table 2,  we summarize and classify exact
solutions obtained above for the model with pp-wave
symmetry.

\begin{tabular}{|l|l|l|}
  \hline
  &  $\eta_{({\rm A})}=0$ & $\eta_{({\rm A})} \neq 0$  \\
\hline
$\phi(u) {=} \phi_0$ & GW of Petrov's type  (\ref{330}) &
Regular periodic  GEMW (\ref{e0358})\\

$\phi_0 \neq 0$ &   &
Regular aperiodic  GEMW (\ref{nqq14}) \\

  &  &
GW with periodic EMW (\ref{Nnm14}) \\

& & Special solution  (\ref{e349}) \\

\hline $\phi^{\prime}(u) \neq 0$ & Symmetric spacetime of GW type
& Axion-GEMW of Petrov's type (\ref{eqq1}) \\

& with axion field linear in time (\ref{nm12}) & Axion-EMW Cheshire smile (\ref{e38}) \\

\hline
 $\phi(u) \equiv 0$& Co-moving GEMW (\ref{nm14}) & Co-moving GEMW
(\ref{nm14}) \\

\hline
\end{tabular}

\vspace{3mm} \noindent {\small Table 2. All the obtained
solutions are classified with respect to the pseudoscalar field
type: $\phi$ is a non-vanishing constant $\phi_0$; $\phi$ depends on
retarded time and $\phi^{\prime}(u)\neq 0$; $\phi$ vanishes (see
horizontal lines); as well as, with respect to values of the
guiding parameter: $\eta_{({\rm A})}{=}0$ or $\eta_{({\rm A})}
\neq 0 $. We distinguish solutions of pure gravitational wave type
(GW), co-moving gravito-electromagnetic waves (GEMW), as well as
axion-GEMW and axion-EMW. The cases indicated as the solutions for
symmetric spacetime relate to the model with covariantly constant
Riemann tensor, the solutions of the Petrov type being the
particular subcase of such spacetime with vanishing Ricci tensor.
In parentheses we indicate the initial formula in the text related
to the corresponding solution. The solutions with $\phi(u) \equiv 0$ are
not new; they are introduced to complete the table. The solutions with
$\eta_{({\rm A})}{=}0$ have, formally speaking, the same structure as
the known minimal solutions for the true scalar field; we focus our attention
to these solutions since they are also valid for the pseudoscalar field with
arbitrary non-vanishing non-minimal coupling parameters $q_1$, $q_2$, $q_3$, $Q_1$ $Q_2$, $Q_3$,
$\eta_1$, $\eta_2$, $\eta_3$, thus belonging to the class of non-minimal solutions.}


\section{Propagation of electromagnetic waves \\ coupled to  the background axion
and \\ gravitational pp-wave fields}

\subsection{Evolutionary equations for the potentials }

Consider now {\it test} electromagnetic waves propagating in an
arbitrary direction in the pure pp-wave background with
$\gamma(u)=0$. The test electromagnetic field satisfies the
non-minimal evolutionary equations (\ref{max2}), (\ref{inducnm})
with a vanishing current on the right-hand side, and the solutions,
describing the axion field $\phi(u)$ and pp-wave gravitational
field, are assumed to be unperturbed. We deal below with pure
gravitational waves; thus, the Ricci scalar and the Ricci tensor
vanish ($R=0$, $R_{uu}=0$), providing that the nonvanishing
components of the Riemann tensor are of opposite signs ($R^2_{\
u2u}=-R^3_{\ u3u}$). Also, we require the potential four-vector
to satisfy the Lorentz gauge condition
\begin{equation}
\nabla_k A^k =0 \ \ \ \Rightarrow \ \ \ \partial_v A_u {+}
\partial_u A_v {+} g^{\alpha \beta} \partial_{\alpha} A_{\beta}
{+} \frac{2L^{\prime}}{L} A_v {=} 0 \,. \label{t1}
\end{equation}
Let us mention that the problem of propagation of electromagnetic
waves coupled to axions in the geometric optics approximation is
studied in the paper \cite{Itin} on the basis of investigation of
the Fresnel equation. Here we do not restrict ourselves by this
approximation and consider exact solutions of the equations of
axion electrodynamics in analogy with solutions of the model of
non-minimal optical activity discussed in \cite{BL1}.

Due to condition (\ref{t1}) the first equation from
(\ref{max2}) (with $i=u$)  accounting for (\ref{inducnm}) gives
the following simple equation for the longitudinal potential $A_v
= \xi^i_{(v)} A_i$:
\begin{equation}
\hat{{\cal D}} A_v  = - \frac{2L^{\prime}}{L}
\partial_v  A_v  \,,
 \label{v}
\end{equation}
where $\hat{{\cal D}}$ is the differential operator:
\begin{equation}
\hat{{\cal D}} \equiv  g^{mn} \partial_m \partial_n = 2 \partial_u
\partial_v {-}L^{-2}\left(e^{-2\beta}\partial^2_2 +
e^{2\beta}\partial^2_3 \right) \,.
 \label{v1}
 \end{equation}
Clearly, (\ref{v}) is an equation for $A_v$ only; it does not
include other components of the potential four-vector. The exact
solution of this equation takes the form (see \cite{1997})
\begin{equation}
A_v  = L^{-1}(u) \ B_v(W)  \,,
 \label{v2}
\end{equation}
where $B_v(W)$ is an arbitrary function of the phase scalar $W$,
given by
\begin{equation}
W \equiv W_0 + \frac{1}{2k_v}\left[k^2_2 \int du L^{-2}e^{-2\beta}
+ k^2_3 \int du L^{-2}e^{2\beta} \right] {+} k_v v  {+} k_2 x^2
{+} k_3 x^3 \,, \label{6}
\end{equation}
with arbitrary constant values of $W_0$, $k_2$, $k_3$ and $k_v$.
The wave four-vector $K_i$
\begin{equation}
K_i \equiv \nabla_i W = - \delta_i^u \frac{1}{2k_v}\left(
g^{22}k_2^2 + g^{33}k_3^2\right) + \delta_i^v k_v + \delta_i^2 k_2
+ \delta_i^3 k_3 \,, \label{1236}
\end{equation}
is the null four-vector, i.e. $g^{im} K_i K_m =0$ and the
longitudinal component of the electromagnetic potential $A_v$
propagates with the speed of light in vacuum. A special solution
$B_v(W)=0$ has been motivated in \cite{1997}. This solution
relates to the Landau gauge condition $ \xi^i_{(v)} A_i =0$, and
we will use it below. Equations for the transversal components
$A_2$ and $A_3$ form a coupled system
\begin{equation}
\left[ \hat{{\cal D}} + 2q_3 R^2_{\ u2u}\partial^2_v -
2\beta^{\prime}
\partial_v \right] A_2 + e^{2\beta}\left[ \phi^{\prime}\partial_v + 2Q_3  \phi
R^2_{\ u2u}\partial^2_v \right] A_3 = 0 \,, \label{2}
\end{equation}
\begin{equation}
\left[ \hat{{\cal D}} + 2q_3 R^3_{\ u3u}\partial^2_v +
2\beta^{\prime}
\partial_v \right] A_3 -
e^{-2\beta}\left[ \phi^{\prime}\partial_v + 2Q_3 \phi R^3_{\
u3u}\partial^2_v \right] A_2 = 0 \,. \label{3}
\end{equation}
The replacement
\begin{equation}
A_2 = e^{\beta} B_2 \,, \quad  A_3 = e^{-\beta} B_3  \label{31}
\end{equation}
simplifies these equations yielding
\begin{equation}
\left[ \hat{{\cal D}} + 2q_3 R^2_{\ u2u}\partial^2_v  \right] B_2
+ \left[ \phi^{\prime}\partial_v + 2Q_3 \phi R^2_{\
u2u}\partial^2_v \right] B_3 = 0 \,, \label{2r}
\end{equation}
\begin{equation}
\left[ \hat{{\cal D}} + 2q_3 R^3_{\ u3u}\partial^2_v  \right] B_3
- \left[ \phi^{\prime}\partial_v + 2Q_3 \phi R^3_{\
u3u}\partial^2_v \right] B_2 = 0 \,. \label{3r}
\end{equation}
Below we obtain and discuss exact solutions of these equations,
suggesting that the functions $\phi(u)$, $L(u)$ and $\beta(u)$ are
presented by one of the exact solutions listed in the previous Section.

\subsection{Birefringence induced by curvature  in the case \\
of vanishing pseudoscalar field ($\phi = 0$) }

In order to clarify a physical meaning of the parameter $q_3$, let
us consider the simplest model with $\phi=0$, i.e. the model
without axion-photon interaction. As it was shown in \cite{1997},
the solution of (\ref{2}) and (\ref{3}) at $\phi = 0$ are
\begin{equation}
A_{2} = e^{\beta}{\cal B}_{(2)}(W_{(2)}) \,, \quad A_3 =
e^{-\beta}{\cal B}_{(3)}(W_{(3)}) \,,  \label{9710}
\end{equation}
\begin{equation}
A_v = 0 \,, \quad A_u = \frac{1}{k_v L^2}\left[k_2 e^{-\beta}
{\cal B}_{(2)}(W_{(2)})  + k_3 e^{\beta} {\cal B}_{(3)}(W_{(3)})
\right] \,. \label{9711}
\end{equation}
Here ${\cal B}_{(2)}$ and ${\cal B}_{(3)}$ are arbitrary functions
of their arguments, $W_{(2)}$ and $W_{(3)}$, given by
\begin{equation}
W_{(2)} = W - q_3 k_v \int_0^u d\tilde{u} R^2_{\
u2u}(\tilde{u})\,, \quad W_{(3)} = W - q_3 k_v \int_0^u d\tilde{u}
R^3_{\ u3u}(\tilde{u}) \,, \label{972}
\end{equation}
with the phase $W$ found in (\ref{6}). The numbers (2) and (3)
indicate here the electromagnetic waves with polarization vectors
directed along $Ox^2$ and $Ox^3$ axes, respectively. For
estimations it is convenient to consider a particular case with
harmonic functions
\begin{equation}
{\cal B}_{(2)} = B_{(2)}^0 \cos{W_{(2)}} \,, \quad {\cal B}_{(3)}
= B_{(3)}^0 \cos{W_{(3)}} \,,  \label{971}
\end{equation}
and parameter $k_v$
\begin{equation}
k_v =  \frac{\omega_{({\rm EM})}}{c\sqrt2} \ (1{-}\cos{\theta})
\label{rotat8}
\end{equation}
explicitly expressed in terms of the electromagnetic wave
frequency $\omega_{({\rm EM})}$ and angle $\theta$ between the
directions of propagation of the gravitational wave ($0x^1$ in our
case) and of the electromagnetic one.

Clearly, $W_{(2)} \neq W_{(3)}$, when the corresponding components
of the curvature tensor do not coincide, $R^2_{\ u2u} \neq R^3_{\
u3u}$, and thus we deal with the birefringence effect induced by
the curvature. The wave four-vectors $K_i^{(2)} = \nabla_i W_{(2)}$
and $K_i^{(3)} = \nabla_i W_{(3)}$ can be represented as follows:
\begin{equation}
K_i^{(2)} = K_i - \delta_i^u q_3 k_v R^2_{\ u2u} \,, \quad
K_i^{(3)} = K_i - \delta_i^u q_3 k_v R^3_{\ u3u} \,, \label{KKK}
\end{equation}
where $K_i$ is given by (\ref{1236}). The wave four-vectors
$K_i^{(2)}$ and $K_i^{(3)}$ are not longer null four-vectors,
since
\begin{equation}
g^{il} K_i^{(2)} K_l^{(2)}=  - 2 q_3 k^2_v R^2_{\ u2u} \,, \quad
g^{il} K_i^{(3)} K_l^{(3)}=  - 2 q_3 k^2_v R^3_{\ u3u}\,.
\label{KKK1}
\end{equation}
The frequencies of the waves with polarization along $Ox^2$ and
$Ox^3$, respectively, (see, e.g., \cite{Cher} for details), are
$$
\omega_{(2)} = \frac{ck_v}{\sqrt2}\left[1 -
\frac{1}{2k_v^2}\left(g^{22}k_2^2 + g^{33}k_3^2 \right) - q_3
R^2_{\ u2u} \right] \,,
$$
\begin{equation}
\omega_{(3)} =
\frac{ck_v}{\sqrt2}\left[1 - \frac{1}{2k_v^2}\left(g^{22}k_2^2 +
g^{33}k_3^2 \right) - q_3 R^3_{\ u3u} \right] \,. \label{FFF}
\end{equation}
These frequencies differ when $R^2_{\ u3u} \neq R^3_{\ u3u}$. The
refraction indices $n_{(2)}$ and $n_{(3)}$ can be calculated as
follows (see, e.g., \cite{Cher} for details):
$$
n^2_{(2)} \equiv 1 - \left(\frac{c}{\omega_{(2)}} \right)^2
g^{il}K^{(2)}_{i}K^{(2)}_{l} = 1 + 2q_3
\left(\frac{ck_v}{\omega_{(2)}} \right)^2 R^2_{\ u2u} \,,
$$
\begin{equation} n^2_{(3)} \equiv 1 -
\left(\frac{c}{\omega_{(3)}} \right)^2
g^{il}K^{(3)}_{i}K^{(3)}_{l} = 1 + 2q_3
\left(\frac{ck_v}{\omega_{(3)}} \right)^2 R^3_{\ u3u} \,.
\label{RRR2}
\end{equation}
The difference $\Delta n \equiv n_{(2)} {-} n_{(3)}$ can be
calculated for all model mentioned in the previous section, but we
focus now on one example only. Let the test electromagnetic
wave propagate in the direction $-x^1$, and the solution for the
gravitational field be of the Petrov type, i.e. $R^2_{\ u2u} = - R^3_{\ u3u} =
\nu^2$. Then one obtains the explicit expressions for the phase
velocities:
\begin{equation}
\frac{v^{(2)}_{\rm ph}}{c} = \frac{\omega_{(2)}}{kc} = 1 - q_3
\nu^2 \,, \quad \frac{v^{(3)}_{\rm ph}}{c} =
\frac{\omega_{(3)}}{kc} = 1 + q_3 \nu^2 \,, \quad k \equiv
\frac{k_v}{\sqrt2} \,, \label{RRR3}
\end{equation}
and for the refraction indices:
\begin{equation}
n_{(2)}  = \frac{1 + q_3 \nu^2}{1 - q_3 \nu^2} \,, \quad n_{(3)} =
\frac{1 - q_3 \nu^2}{1 + q_3 \nu^2} \,,  \quad \Delta n =
\frac{4q_3 \nu^2}{1- q_3^2 \nu^4} \,. \label{RRR4}
\end{equation}
This means, first, that the phase velocities differ from each
other, one of them being less than the speed of light in vacuum $c$,
while the second being bigger than $c$. Second, the non-minimal
(curvature induced) difference between the refraction indices for
the waves with orthogonal polarizations depends on the curvature
$\nu^2$ and the non-minimal coupling parameter $q_3$. Thus, in
this illustration, we deal with birefringence in the absence of
polarization rotation.

The constraints on the non-minimal coupling parameter from solar
system observations and from pulsar data analysis were studied in
\cite{PM} (the parameter $\lambda$ in this paper relates to
$q_3/4$ in our case). From the timing observations of the binary
pulsar PSR B1534+12 signals, Prasanna and Mohanty gave a
constraint on $q_3$ to be $|q_3| < 2.4 \cdot 10^{11} \ {\rm
cm}^2$. From more precise observations of the double pulsar system
PSR J0737-3039 A/B, the Shapiro time delay (passing PSR J0737-3039
B of mass 1.25 $M_{\rm Sun}$ and radius 10 km) is measured to be
$6.21 \pm 0.33 \ {\rm \mu s}$ \cite{76a}. The theoretical estimate
using general relativity is $6.153 \pm 0.26 \ {\rm \mu s}$. If we
take an upper limit of $0.1 \ {\rm \mu s}$ for the maximum
contribution of the polarization-dependent correction, we have an
upper bound $|q_3| < 2.4 \cdot 10^{10} \ {\rm cm}^2$.

\subsection{Quasi-minimal model of optical activity
($q_3=0$, $Q_3=0$) \\ induced by varying pseudoscalar field ($\phi
\neq \phi_0$)}

Now we consider the model in which the birefringence is absent,
but the rotation of polarization is  present. Let two non-minimal
coefficients $q_3$ and $Q_3$ be vanishing; then the exact
solutions of (\ref{t1}), (\ref{v}), (\ref{2}), (\ref{3}) are
\begin{equation}
A_2= e^{\beta(u)}\left\{E(W) \cos{\frac{1}{2}[\phi(u){-} \phi(0)]}
+ F(W) \sin{\frac{1}{2}[\phi(u) {-} \phi(0)]} \right\} \,,
\label{4}
\end{equation}
\begin{equation}
A_3= e^{-\beta(u)}\left\{E(W) \sin{\frac{1}{2}[\phi(u) {-}
\phi(0)]} - F(W) \cos{\frac{1}{2}[\phi(u) {-} \phi(0)]} \right\}
\,, \label{4r}
\end{equation}
\begin{equation}
A_v =0 \,, \quad A_u =  \frac{1}{k_v L^2} \left[k_2 e^{-2\beta(u)}
A_2 + k_3 e^{2\beta(u)} A_3 \right]\,. \label{5}
\end{equation}
Here $E(W)$ and $F(W)$ are arbitrary functions of one phase scalar
$W$ (\ref{6}); thus, there is no birefringence. But now we deal
with the effect of Faraday rotation, induced by the photon-axion
interactions. This is a kind of optical activity induced by the axion
field. The effect of the Faraday rotation is described by the rotation
phase $\phi(u){-}\phi(0)$, so this effect disappears when the
pseudoscalar field is constant. When the pseudoscalar field is
a linear function of retarded time (see (\ref{nm12}), (\ref{nm13})),
the Faraday rotation is uniform. Let us mention again that we
indicate this model as quasi-minimal since we suppose that only
two non-minimal parameter are vanishing, but other ones are
non-vanishing.

\subsection{Non-minimal optical activity induced by photon-axion
\\ interaction in the case of a constant pseudoscalar field $\phi=\phi_0$}

When non-minimal coupling constants are non-vanishing, i.e. $q_3
\neq 0$ and $Q_3 \neq 0$, we search for exact solutions to the
system (\ref{2r}), (\ref{3r}) as follows:
\begin{equation}
B_2 =  h_1(u) \cos{W} + h_2(u) \sin{W} \,, \quad B_3 = - h_3(u)
\sin{W} + h_4(u) \cos{W} \,. \label{nm1}
\end{equation}
Again $W$ is given by (\ref{6}), and four functions of the
retarded time $h_1$, $h_2$, $h_3$ and $h_4$ satisfy the following
system of linear ordinary first-order differential equations with
a coefficient depending on retarded time $u$:
$$
h^{\prime}_1(u) = - \frac{1}{2} \phi^{\prime} \ h_4 - q_3  k_v
R^2_{\ u2u} \ h_2 + Q_3  \phi  \ k_v  R^2_{\ u2u} \ h_3 \,,
$$
$$
h^{\prime}_2(u) =  + \frac{1}{2} \phi^{\prime} \ h_3 + q_3  k_v
R^2_{\ u2u} \ h_1 + Q_3  \phi  \ k_v  R^2_{\ u2u} \ h_4 \,,
$$
$$
h^{\prime}_3(u) = - \frac{1}{2} \phi^{\prime} \ h_2 - q_3  k_v
R^3_{\ u3u} \ h_4 + Q_3  \phi  \ k_v  R^3_{\ u3u} \ h_1 \,,
$$
\begin{equation}
h^{\prime}_4(u) = + \frac{1}{2} \phi^{\prime} \ h_1 + q_3  k_v
R^3_{\ u3u} \ h_3 + Q_3  \phi  \ k_v  R^3_{\ u3u} \ h_2 \,.
\label{nm2}
\end{equation}
Let the function $\phi(u)$ be presented by a constant value $\phi(u)
= \phi_0$. Introducing a new convenient variable $\tau$
\begin{equation}
\tau \equiv k_v \int_0^u  R^2_{\ u2u} (\tilde{u})d\tilde{u} \,,
\label{nm3}
\end{equation}
we reduce the system (\ref{nm2}) to the following system with
constant coefficients:
$$
\dot{h}_1(\tau) =  - q_3 h_2 + Q_3 \phi_0  h_3 \,, \quad
\dot{h}_2(\tau) =   q_3  h_1 + Q_3 \phi_0  h_4 \,,
$$
\begin{equation}
 \dot{h}_3(\tau) =  - Q_3 \phi_0  h_1  + q_3  h_4  \,, \quad
\dot{h}_4(\tau) =  - Q_3 \phi_0  h_2 - q_3  h_3  \,, \label{nm5}
\end{equation}
where the dot denotes the derivative with respect to $\tau$.
The characteristic equation for this system of equations with
constant coefficients takes the form
\begin{equation}
\left(\lambda^2 + Q^2_3 \phi_0^2 + q_3^2 \right)^2 = 0 \,,
\label{nm81}
\end{equation}
and its solutions $\lambda = \pm i \Omega$ are the double ones,
where
\begin{equation}
\Omega \equiv \sqrt{q_3^2 + Q_3^2 \phi_0^2} \,. \label{nm810}
\end{equation}
Let us mention that (\ref{nm5}) in this case can be rewritten in
terms of second-order equations:
\begin{equation}
\ddot{h}_{a} + \Omega^2 h_{a} = 0 \,, \label{nm81r}
\end{equation}
where $a=1,2,3,4$. That is why the exact solution of the system
(\ref{nm5})
$$
h_1(\tau) = h_1(0) \cos{\Omega \tau} - \frac{1}{\Omega}
\sin{\Omega \tau} [ q_3 \ h_2(0) - Q_3 \phi_0 \ h_3(0) ] \,,
$$
$$
h_2(\tau) =  h_2(0) \cos{\Omega \tau} + \frac{1}{\Omega}
\sin{\Omega \tau} [ q_3  \ h_1(0) + Q_3 \phi_0 \ h_4(0) ] \,,
$$
$$
h_3(\tau) =  h_3(0) \cos{\Omega \tau} + \frac{1}{\Omega}
\sin{\Omega \tau} [ q_3 \ h_4(0) - Q_3 \phi_0 \ h_1(0) ] \,,
$$
\begin{equation}
h_4(\tau) = h_4(0) \cos{\Omega \tau} - \frac{1}{\Omega}
\sin{\Omega \tau} [q_3 \ h_3(0) + Q_3 \phi_0  \ h_2(0) ] \,,
\label{nm6}
\end{equation}
contains the trigonometric functions $\cos{\Omega \tau}$ and
$\sin{\Omega \tau}$ only. Thus, the quantity $\Omega$ plays a role
of frequency of the curvature-induced rotation of the Faraday
type. Let us stress that in this model the non-minimal coupling
parameters $Q_3$ and $q_3$ play equivalent roles, predetermining
the pure rotation of the polarization vector without damping. When
the pp-wave spacetime is symmetric, i.e. the component $R^2_{\
u2u}$ is constant, one obtains immediately from (\ref{nm3}) that
$\tau$ is linear in the retarded time, e.g. for solution
(\ref{330}) one obtains $\tau = \nu^2 k_v u$.

\subsubsection{Constraints on the non-minimal coupling parameters
from CMB polarization observations}

The parameter $Q_3$ enters the key formulas in the product $Q_3
\phi_0$ and thus relates to the non-minimal axion contribution to
the effect of polarization rotation. Let us assume that $q_3 = 0$
and the following simple initial conditions take place
\begin{equation}
h_1(0) = h_3(0) = h_4(0)=0 \,, \quad h_2(0) = B_0  \ \ \rightarrow
\ \  B_2 = B_0 \ \sin{W} \,, \quad B_3 = 0 \,, \label{rotat4}
\end{equation}
i.e. initially the electromagnetic wave was polarized along the
$0x^2$ axis only. Then we obtain from (\ref{nm6})
\begin{equation}
A_2 = B_0 e^{\beta} \cos{(Q_3\phi_0 \tau)}  \sin{W} \,, \quad A_3
= - B_0 e^{-\beta}  \sin{(Q_3\phi_0 \tau)} \cos{W} \,.
\label{rotat5}
\end{equation}
Clearly, when $Q_3\phi_0 \neq 0$, the orthogonal component of the
electromagnetic potential $A_3$ appears;  thus, the polarization
rotation can exist when the parameter $Q_3\phi_0$ is
non-vanishing. The polarization ellipticity can be described in
this case by the formula
\begin{equation}
\frac{\left(A_2e^{-\beta} / B_0\right)^2}{\cos^2{(Q_3\phi_0
\tau)}} + \frac{\left(A_3e^{\beta} /
B_0\right)^2}{\sin^2{(Q_3\phi_0 \tau)}} = 1 \,. \label{rotat6}
\end{equation}
Thus, the polarization rotation rate is predetermined by the
function $Q_3\phi_0 \tau$, and the quantity
\begin{equation}
[\nabla \phi]_{\rm effective} = Q_3 \phi_0 \sqrt2 \
\left[\frac{H_{({\rm GW})}}{\lambda^2_{({\rm GW })}} \right] \
\frac{\omega_{(EM)}}{c} \ (1{-}\cos{\theta})  \label{rotat7}
\end{equation}
can be considered as an equivalent of the gradient of the pseudoscalar
field for the estimates of the effect. The frequency dependence is
different from the Faraday rotation in an ionized medium. From
(\ref{rotat7}), the order of magnitude of  $[\nabla \phi]_{\rm
effective}$ from a microwave of frequency 100 GHz is
\begin{equation}
[\nabla \phi]_{\rm effective} \simeq 4.7 \ Q_3 \phi_0
\left[\frac{H_{({\rm GW})}}{\lambda^2_{({\rm GW })}} \right] {\rm
cm}^{-1} \,, \label{rotat71}
\end{equation}
hence,
\begin{equation}
|\Delta \phi_{\rm effective}| \simeq 4.7 \ {\rm cm}^{-1} Q_3
\phi_0 \left[\frac{H_{({\rm GW})}}{\lambda_{({\rm GW })}} \right]
 \,. \label{rotat711}
\end{equation}
Let $ H_{({\rm GW})} =  10^{-15} \xi$, where $\xi \leq 1$; then
for
\begin{equation}
\lambda_{({\rm GW})} = {\rm Hubble \ Distance} = 1.3 \cdot 10^{28}
{\rm cm}  \label{rota1}
\end{equation}
we have
\begin{equation}
|\Delta \phi_{\rm effective}| \simeq 3.6 \cdot 10^{-43} \xi {\rm
cm}^{-2} Q_3 \phi_0 \,. \label{rota11}
\end{equation}
Constraints on cosmic polarization rotation angle $\varphi$ come
from three kinds of observations: (i) polarization observations of
radio galaxies; (ii) optical/UV polarization observations of radio
galaxies; (iii) CMB polarization observations. Radio observations
put a limit of $\Delta \varphi \leq 0.17 - 1.0 \ {\rm rad}$ over
cosmological distance from various types of analysis (see, e.g.,
\cite{19} for a review). Optical/UV observations put a limit of
$\Delta \varphi \leq 0.17 \ {\rm rad}$ \cite{53,54}. The QUaD
observations \cite{Pryke} on CMB polarization give the most
stringent constraint on $\Delta \varphi$ \cite{Wu}
\begin{equation}
\Delta \varphi = 0.0096 \pm 0.0143 \pm 0.008 \ {\rm rad}  \,.
 \label{estim1}
\end{equation}
For gravitational waves with a wavelength of the order of the Hubble
distance, this is the constraint, $\Delta \varphi_{{\rm HD}}$; for
shorter wavelength, $\lambda$, we have to use the estimate
\begin{equation}
\Delta \varphi_{\lambda} \simeq  \Delta \varphi_{{\rm HD}}
\left\{\frac{\lambda}{{\rm Hubble \ Distance}}\right\} \,.
 \label{estim2}
\end{equation}
In the general case, $\Delta \varphi$ depends on directions
\cite{20}. Constraints on polarization rotation could also come
from solar system observations: the analysis of these data are
under study. Combining the estimate (\ref{rota11}) with
(\ref{estim2}) and using the relation $\phi=2\varphi$, we have the
constraint $Q_3 \phi_0 < 2 \cdot 10^{41} \xi^{-1}{\rm cm}^2$.

\section{Conclusions}

\noindent 1. We formulated a new {\it non-minimal ten-parameter}
Einstein-Maxwell-axion model, i.e. on the basis of the Lagrangian
approach, we derived a non-minimally extended self-consistent
(coupled) system of equations for electromagnetic (see equations
(\ref{max2}),(\ref{inducnm}),(\ref{current})), pseudoscalar (see
equation (\ref{eqaxi1})) and gravitational (see
(\ref{Eineq})-(\ref{calT7})) fields. The Lagrangian is linear in
the curvature tensor and its contractions, and is quadratic in the
Maxwell tensor, i.e. we deal with one of the versions of {\it
non-minimal linear axion electrodynamics}.

\noindent 2. The constitutive tensor (see (\ref{nminduc}) with
(\ref{sus1})-(\ref{sus3})), associated with this model, is
manifestly symmetric with respect to transposition of pairs of
indices, i.e. this model excludes skewons \cite{HehlObukhov} and
describes non-minimal interaction of photons and axions only. An
extension of the model accounting for skewon-type interactions
will be considered separately.

\noindent 3. Decomposition of the constitutive tensor demonstrates
explicitly that the dielectric permittivity, magnetic
impermeability tensors and the tensor of cross-effects (or tensor
of magneto-electric coefficients, in other words) acquire
non-minimal contributions, including terms proportional to the
pseudoscalar (axion) field. Since the tensor of non-minimal
cross-effects is non-vanishing, one may conclude that, generally,
the curvature-induced effect of {\it optical activity} is
expected, even if the pseudoscalar (axion) field is constant.

\noindent 4. We applied the non-minimal Einstein-Maxwell-axion
model to the case when the spacetime has the {\it pp-wave}
symmetry and all the physical fields inherit this symmetry. The
reduced system of equations contains in this case only one
key-equation (\ref{Lpp}). We presented nine examples of exact
solutions for the non-minimally coupled electromagnetic, axion
and gravitational fields with pp-wave symmetry.
One can emphasize three new solutions among them. First,
there is a new {\it regular} solution (see section 3.3.1.), which can
appear in the non-minimal model only; the term regularity means
that gravitational, electromagnetic and pseudoscalar fields have
no singularities and are described by the functions finite
everywhere. Second, there exist exact solutions describing
circularly polarized electromagnetic waves coupled non-minimally
to the axion field and plane gravitational waves (see
(\ref{eqq1})-(\ref{Reqq13})). Third, the solution which is known
as {\it Cheshire smile} is also admissible in this model (see
section 3.3.4).

\noindent 5. Discussing the {\it optical activity} induced by
photon-axion interaction, we solved exactly the problem of
propagation of test electromagnetic waves coupled non-minimally to
the pseudoscalar and gravitational fields in the pp-wave
background. According to the classification of the models of
photon-axion interaction given in the paper \cite{Itin}, we deal
here with the case when the four-gradient of the pseudoscalar
field is a null four-vector, i.e. $\nabla_k \phi \nabla^k \phi
=0$. We show explicitly that the non-minimal coupling of the
photon and axions with gravitational field generally leads to the
{\it birefringence} effect. We discussed explicit exact solutions
which describe the effect of optical activity.

\noindent 6. Non-minimal Einstein-Maxwell-axion model can give the
cosmological solutions of the FLRW and Bianchi-I type, as well as
static spherically symmetric solutions of the
Reissner-Nordstr$\ddot{{\rm o}}$m type. We intend to consider the
corresponding exact solutions in the near future.

\newpage

\noindent
{\bf Acknowledgments}

\noindent
We thank the National Natural Science Foundation of China
(Grant No. 10875171) and Russian Foundation for Basic Research
(Grants No. 08-02-00325-a and 09-05-99015) for support. A.B. is
grateful to colleagues from the Center for Gravitation and
Cosmology of Purple Mountain Observatory of Chinese Academy of
Science for hospitality.

\end{document}